\documentclass[sigconf]{acmart}

\usepackage{amsmath,amssymb,amsfonts}
\usepackage{algorithmic}
\usepackage{graphicx}
\usepackage{textcomp}
\usepackage{makecell}
\usepackage{multirow}
\usepackage{tabularx}
\usepackage{pgfplots}
\usepackage{subcaption}
\usepackage{xspace}

\usepackage[disable]{todonotes}
\usepackage{soul}          
\usepackage{tabularx}      
\usepackage{booktabs}      
\usepackage{enumitem}      
\usepackage{xspace}
\usepackage{listings}
\usepackage{algorithm2e}
\newcommand{\new}[1]{#1}
\newcommand{\old}[1]{}

\copyrightyear{2026}
\acmYear{2026}
\setcopyright{cc}
\setcctype{by-nc-nd}
\acmConference[CHI '26]{Proceedings of the 2026 CHI Conference on Human Factors in Computing Systems}{April 13--17, 2026}{Barcelona, Spain}
\acmBooktitle{Proceedings of the 2026 CHI Conference on Human Factors in Computing Systems (CHI '26), April 13--17, 2026, Barcelona, Spain}
\acmPrice{}
\acmDOI{10.1145/3772318.3790954}
\acmISBN{979-8-4007-2278-3/2026/04}

\begin{document}


\title[Conversational Inoculation]{Conversational Inoculation to Enhance Resistance to Misinformation}

\author{Dániel Szabó}
\orcid{0009-0003-7299-9116}
\email{daniel.szabo@oulu.fi}
\affiliation{%
  \institution{University of Oulu}
  \streetaddress{Pentti Kaiteran katu 1}
  \city{Oulu}
  \country{Finland}
  \postcode{90570}
}

\author{Chi-Lan Yang}
\email{chilan.yang@iii.u-tokyo.ac.jp}
\orcid{0000-0003-0603-2807}
\affiliation{%
  \institution{The University of Tokyo}
  \city{Tokyo}
  \country{Japan}
  \postcode{113-0033}
}

\author{Aku Visuri}
\email{aku.visuri@oulu.fi}
\orcid{0000-0001-7127-4031}
\affiliation{%
  \institution{University of Oulu}
  \streetaddress{Pentti Kaiteran katu 1}
  \city{Oulu}
  \country{Finland}
  \postcode{90570}
}

\author{Jonas Oppenlaender}
\email{jonas.oppenlaender@oulu.fi}
\orcid{0000-0002-2342-1540}
\affiliation{%
  \institution{University of Oulu}
  \streetaddress{Pentti Kaiteran katu 1}
  \city{Oulu}
  \country{Finland}
  \postcode{90570}
}

\author{Bharathi Sekar}
\email{bharathi.sekar@oulu.fi}
\orcid{0009-0006-1421-5063}
\affiliation{%
  \institution{University of Oulu}
  \streetaddress{Pentti Kaiteran katu 1}
  \city{Oulu}
  \country{Finland}
  \postcode{90570}
}

\author{Koji Yatani}
\email{koji@iis-lab.org}
\orcid{orcid=0000-0003-4192-0420}
\affiliation{%
  \institution{The University of Tokyo}
  \city{Tokyo}
  \country{Japan}
  \postcode{113-0033}
}

\author{Simo Hosio}
\orcid{0000-0002-9609-0965}
\email{simo.hosio@oulu.fi}
\affiliation{%
  \institution{University of Oulu}
  \streetaddress{Pentti Kaiteran katu 1}
  \city{Oulu}
  \country{Finland}
  \postcode{90570}
}

\renewcommand{\shortauthors}{Szabó et al.}
\renewcommand{\sectionautorefname}{Section}
\renewcommand{\subsectionautorefname}{Section}
\renewcommand{\subsubsectionautorefname}{Section}

\begin{abstract}


Proliferation of misinformation is a globally acknowledged problem. Cognitive Inoculation helps build resistance to different forms of persuasion, such as misinformation. We investigate Conversational Inoculation, a method to help people build resistance to misinformation through dynamic conversations with a chatbot. We built a Web-based system to implement the method, and conducted a within-subject user experiment to compare it with two traditional inoculation methods. Our results validate Conversational Inoculation as a viable novel method, and show how it was able to enhance participants' resistance to misinformation. \old{A qualitative analysis of the conversations between participants and the chatbot revealed independence and trust as factors that boosted the efficiency of Conversational Inoculation, and friction of interaction as a factor hindering it.}\new{A qualitative analysis of the conversations between participants and the chatbot highlighted adaptability, independence, trust and friction as the main factors affecting Conversational Inoculation.} We discuss the opportunities and challenges of using Conversational Inoculation to combat misinformation. Our work contributes a timely investigation and a promising research direction in scalable ways to combat misinformation.



\end{abstract}


\newcommand{\simo}[1]{\textcolor{blue}{}}
\newcommand{\dani}[1]{\textcolor{red}{}}




\begin{teaserfigure}
\centering
  \includegraphics[width=0.9\linewidth]{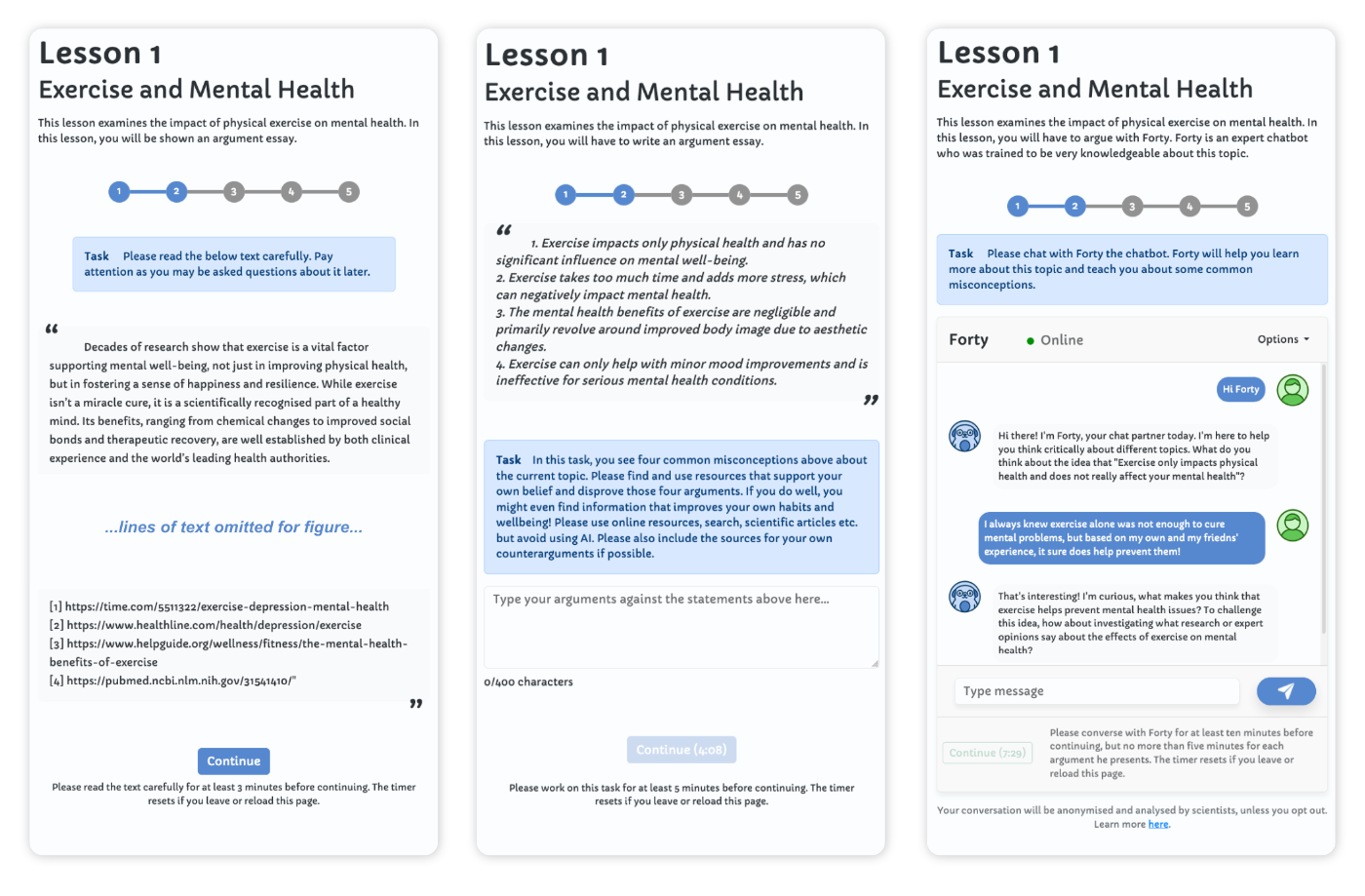}
  \caption{The MindFort system implemented three experimental conditions (the treatments). From left: (a) Reading, (b) Writing, and (c) Chatbot condition. The Chatbot condition yielded greater resistance to misinformation than Reading or Writing, after controlling for baseline susceptibility.}
  \Description{A side-by-side comparison of the user interfaces for the three experimental conditions in Lesson 1: Exercise and Mental Health. Left Panel: Reading Condition. The screen displays an essay about the mental health benefits of exercise, with a task box instructing the participant to read carefully and be prepared for questions. Middle Panel: Writing Condition. The screen shows four common misconceptions about exercise, followed by a task box that asks the participant to research and write counterarguments in a text box. Right Panel: Chatbot Condition. The screen shows a conversational interface with a chatbot named Forty. The chatbot prompts the participant with a misconception about exercise and mental health, and the participant replies in the chat box. Forty responds with follow-up questions, encouraging critical thinking. A progress bar at the top of each screen indicates lesson progress through five steps.}
  \label{fig:herocollage}
\end{teaserfigure}

\maketitle

\section{Introduction}

Misinformation is now a global problem. 
For instance, false information that distorts evidence on nutrition, vaccines, and treatments poses a direct threat to public health. 
False claims about nutrition, vaccines, and treatments threaten public health by encouraging unhealthy habits~\cite{swire2020public} and fostering misconceptions that lead to poor decisions on issues such as vaccination and cancer care.
The COVID-19 pandemic offered a stark example, as widespread misinformation contributed to the loss of lives worldwide \cite{FerreiraCaceres2022}.
Despite progress in combating misinformation through both content-centred methods (e.g., detection) \cite{junhao25} and human-centred strategies (e.g., education) \cite{junhao24}, the rise of generative AI has made misinformation more credible and easier to produce than ever before \cite{Zhang_Sharma_Du_Liu_2024}.


Cognitive Inoculation Theory (CIT)~\cite{McGuire_1961,Banas_Rains_2010}, developed in the early 1960s, conceptualises \textit{cognitive inoculation} as a systematic method for strengthening resistance to persuasion.
Originally applied to attitude change, the framework has since been validated across domains and is increasingly relevant to contemporary challenges such as online misinformation.
Cognitive inoculation has a simple biological metaphor at its core: Much like the human immune system can be prepared with weakened forms of viruses via vaccines, the mind can develop resistance to misinformation via strategic and facilitated exposure to weakened forms of misinformation \cite{Banas_Rains_2010}.
CIT 
is well-documented over decades of research, along with a notable increase in research interest over the past two decades \cite{Compton_2024, Banas_Rains_2010}. 
However, traditional inoculation methods often do not adapt to an individual's pre-existing attitudes or reactions during the intervention itself \cite{Zarzosa_Ruvalcaba_2025}.
Recent research has identified the potential of Large Language Models (LLMs), particularly through chatbots, to support a wide range of health communication tasks
~\cite{Passanante_Pertwee_Lin_Lee_Wu_Larson_2023, wei2024leveraging}.
Yet, cognitive inoculation via conversational interfaces such as LLM-powered chatbots remains largely underexplored.

In this paper, we introduce the concept of \textit{Conversational Inoculation}, a paradigm in which participants' resistance to persuasion is developed through structured and interactive conversation with chatbots. 
We developed \textit{MindFort}: a Web-based system for Conversational Inoculation. 
MindFort implements an LLM-powered \textit{chatbot} for inoculation.
For the purpose of the experiment, we also implemented a \textit{reading} condition (representing a supportive defense treatment in which participants are passively educated about the subject) and a \textit{writing} condition (a traditional cognitive inoculation approach involving research and refutation).
In conversational systems such as our LLM-powered chatbot, and in real life, conversation can take place in many forms, where dynamics are influenced by the number of parties, their roles and relationships.
In our study, MindFort implements Conversational Inoculation through a structured, real-time dialogue with a chatbot named Forty. The chatbot assumes the role of a proactive debate partner, intentionally guiding the user through a process of developing and rehearsing refutations against persuasion.
This approach addresses the main limitations of traditional inoculation methods. MindFort is an interactive system capable of adapting to the user's engagement, views, and experiences, and encourages them to practice critical thinking and research skills in the process \cite{Zarzosa_Ruvalcaba_2025}.
Our work focuses on health-related information, as a common \cite{Banas_Rains_2010}, relevant \cite{Passanante_Pertwee_Lin_Lee_Wu_Larson_2023}, and an impactful domain where advances in communication and education yield direct and significant benefits for individuals and communities \cite{Compton_Jackson_Dimmock_2016}.
Cognitive Inoculation, in particular, can reinforce healthy, fact-based beliefs to promote reliance on evidence, protect healthy attitudes and behaviours, and ultimately support public wellbeing  \cite{Compton_Jackson_Dimmock_2016}.

We evaluated MindFort in an online within-subject experiment with 65 participants to investigate how conversational inoculation compares against the two alternative treatments of reading and writing. 
Our work addresses the following three research questions:

\textbf{RQ1:} How effective is conversational inoculation, as implemented in MindFort, in enhancing resistance to misinformation compared to traditional approaches of reading and writing?




\textbf{RQ2:} How do participants' experience with conversational inoculation compare to traditional approaches of reading and writing?




\old{\textbf{RQ3:} What linguistic features and mechanisms act as promotors or inhibitors of Conversational Inoculation?}
\new{\textbf{RQ3:} What linguistic features and cognitive mechanisms are associated with Conversational Inoculation?}



Our study empirically validates Conversational Inoculation as an effective method for improving resistance to misinformation, although its advantage over traditional methods is nuanced. 
Crucially, we found that the conversational format is engaging and can be implemented without sacrificing user motivation.
We discuss the future potential of Conversational Inoculation, arguing that its key benefits are not just the conversational format itself but also its ability to facilitate an inoculation intervention in ways traditional methods cannot.
Therefore, our work contributes to HCI by solidifying Conversational Inoculation as a promising research direction in the important fight against misinformation.

\section{Background}

Our study is best contextualised within the Cognitive Inoculation Theory, inoculation in the field of HCI, and how conversational techniques can combat misinformation, particularly in health-related topics. 

\subsection{Cognitive Inoculation Theory}

Cognitive Inoculation theory (often just referred to as Inoculation Theory) originates from social psychology.
The theory posits that exposure to a weakened form of an argument can strengthen resistance to subsequent persuasion attempts \cite{Compton_2024, McGuire_1961, McGuire_1964}.
A useful analogy is medical vaccination: just as vaccines build immunity to disease, preemptive exposure to counterarguments and their refutations can strengthen resistance against subsequent persuasion attempts.

\citet{McGuire_1961} showed that pre-exposing participants to weakened arguments against an attitude or belief they currently hold, known as \textit{inoculation}, makes them more resistant to subsequent persuasion.
Specifically, participants who received inoculation were less affected by a subsequent strong counterattitudinal (going against one's current attitude) message than those who were either shown only supporting statements or given no treatment at all. Inoculation Theory has been used for example, to reduce susceptibility to misinformation \cite{Roozenbeek_VanDerLinden_Nygren_2020}, politics \cite{Pfau1988INOCULATIONIP}, commerce \cite{inoculation_advertising_2016}, and --- of particular importance to our study --- health \cite{compton2017inoculation, McGuire_1961, pfau1992use}. As shown by \citeauthor{Compton_2024}'s recent review, Inoculation Theory is increasingly researched: in 2023 alone, 450 publications indexed by Google Scholar mentioned Inoculation Theory, whereas this number was 132 in 2013 and 61 in 2003, and there are meta-analyses offering empirical support for the theory \cite{Banas_Rains_2010}. 

In their meta-analysis in 2010, \citet{Banas_Rains_2010} reviewed 50 years of Inoculation Theory research. After the original work \cite{McGuire_1961, McGuire_1964} by \citeauthor{McGuire_1961}, this review verified the core prediction of Inoculation Theory. Further, \citeauthor{Banas_Rains_2010} point out moderators of inoculation effectiveness such as the level of perceived threat or the participant's issue involvement, which may inform design decisions of our system. In their recent work, \citeauthor{Fransen_Mollen_Rains_Das_Vermeulen_2024} replicated \citeauthor{McGuire_1961}'s 1961 experiments with minor changes to the experimental design and contents to bridge the cultural gap of 60 years.

As evident from \citet{Roozenbeek_VanDerLinden_2019}'s work on a gamified, interactive web-based cognitive inoculation delivery system, the \textit{Bad News} game can inoculate people against misinformation with success. Their system differs a lot from traditional, less interactive inoculation techniques and suggests that this new direction is effective at increasing people's resistence to persuasion, especially in a time when attention spans are shorter and there are many things competing for our attention at all time. The game \textit{Bad News} is very focused on political misinformation spread on short-form text based social media (e.g. Twitter), and the great results achieved suggest that a more broad, conversational approach should be investigated.

\subsection{Inoculation in HCI}

Recognizing the inadequacy of purely technical solutions for combating misinformation, Human-Computer Interaction (HCI) researchers are increasingly exploring human-centered, behavioral interventions \cite{Konstantinou_Karapanos_2025}. Among these, Cognitive Inoculation (CI) theory is gaining traction as a promising approach. Both recent scoping reviews and surveys highlight inoculation as a valid method, while also noting that technology-based implementations are still in an early, experimental stage \cite{Konstantinou_Karapanos_2025, Fard_Lingeswaran_2020}.

\citeauthor{Barman_Conlan_2024} confirmed the effectiveness of prebunking, but their results highlighted the necessity of user-centric approaches and tailored interventions. 
Gamified approaches have also shown promising outcomes in inoculation against misinformation\cite{Basol_Roozenbeek_VanDerLinden_2020}, deceptive e-commerce practices \cite{Aung_Soubutts_Singh_2024} and echo chamber effect on social networking services \cite{Jeon_Kim_Xiong_Lee_Han_2021}. \citeauthor{Tang_Sergeeva_2025} advanced the gamified approach to inoculation against misinformation as they integrated LLMs in their game where the player learns misinformation tactics by trying to convince a simulated public of LLM agents \cite{Tang_Sergeeva_2025}. 
The system was highly engaging, immersive and successful in enhancing subjects' ability to recognise misinformation and even counter it. Building on the strengths of this study, \citet{Tang_Sergeeva_2025} presented a design concept for a mobile game, which includes an exchange between the user and an AI assistant regarding a presented piece of misinformation related to a fictional pandemic similar to COVID-19. 
In \citeauthor{Tang_Sergeeva_2025}'s design the AI assistant provides personalised feedback to the user's reaction to the misinformation presented to them, and the system is envisioned to tackle visual as well as textual misinformation, and cover a large variety of topics.

A typical limitation of prior inoculation techniques is that the inoculation process does not consider the pre-inoculation attitude of the subject \cite{Zarzosa_Ruvalcaba_2025} or the subject's reaction to the inoculation message.
The modern capabilities of LLM-powered conversational systems are uniquely suitable for tailoring messages to the user and, for instance, changing the tone and approach during the process.

\subsection{Conversational Solutions for Combating Misinformation}

Novel methods approaching misinformation resilience through conversational techniques have gained attention lately.
Several conversational approaches, in particular chatbots (used often interchangeably with conversational agents), have been developed to help users combat misinformation, learn about misinformation, and become more resistant to it.
For instance, recent work has explored deploying fact-checking capabilities directly within messaging platforms, e.g. \textit{Auntie Meiyu}, a chatbot integrated into LINE messaging to help users protect family members from misleading news in private messaging groups~\cite{Lee2025}. 
Their evaluation revealed mixed user reactions. While users appreciated the bot's protective intent, they expressed concerns about its robotic communication style and the occasional errors. 

The success of conversational anti-misinformation tools depends on user trust and sustained engagement. 
\citet{Rossner2025263} examined user frustration in AI-driven chatbot interactions, emphasising that accuracy and transparency are crucial for maintaining user engagement in misinformation contexts. 
Understanding user motivations is equally critical. 
Research has shown how both situational and gratification motivations significantly contribute to users' willingness to engage in communicative actions that aim to reduce misinformation spread~\cite{Cheng2025643}.

The healthcare domain has seen particular attention in conversational misinformation research.
Interactive and empathetic conversational design features influence health misinformation correction and vaccination intentions~\cite{Gong2025276}: interactive chatbots were associated with lower levels of health misperceptions, while empathetic chatbots directly increased users' vaccination intentions. 
Vaccines, in particular, have been a prominent topic in misinformation research with chatbots. 
In a 2023 systematic review, \citet{Passanante_Pertwee_Lin_Lee_Wu_Larson_2023} examined documented cases of use of conversational systems for vaccine communication, as vaccine misinformation and vaccine hesitation were a focal point during the recent COVID-19 pandemic.
The seven identified articles, all published before August 2022, employed either rule-based systems or Wizard of Oz experiments.
The authors found evidence of potential benefits of the conversational approach for vaccine communication.

In their 2024 article, \citet{Peng_Lee_Lim_2024} studied conversational agents during the COVID-19 pandemic, focusing on chatbots as tools for education and interaction with online content.
Through participatory design workshops with older adults, they identified design implications for conversational systems aimed at countering misinformation, such as using well-trusted sources, practicing transparency, and the need to go beyond just fact-checking.

Further, \citeauthor{Fasce_2025}'s 2025 work \cite{Fasce_2025} shows that conversational approaches to improving health attitudes are effective.
The authors evaluated two conversational techniques, Empathetic Refutational Interviewing and Motivational Interviewing to address vaccine hesitancy among patients in Romania and found that patients' attitude and willingness to get vaccinated have improved.

In 2025 Large Language Models have already been employed as a tool to make cognitive inoculation more accessible. \citet{Malek_2025} addressed one of the main limitations of CIT, the effort required to develop materials for each subject.
The authors identified that LLMs can assist in preparing the inoculation materials in their case study of COVID-19-related online misinformation, going as far as identifying the misinformation intelligently and deploying the refutations automatically.

\section{Method}

We explore the gap in the intersection of the discussed related work areas, namely, employing CIT through a conversational system.
We developed a custom Web-based system, MindfFort, for the inoculation experiment.
All source code \footnote{\url{https://github.com/Crowd-Computing-Oulu/mindfort}} and data of 65 participants \footnote{\url{https://github.com/Crowd-Computing-Oulu/conversational-inoculation-study-data-2026}}
are openly available.


\subsection{Ethics}
The study protocol underwent review by the departmental Institutional Review Board (IRB) and received a full review and ethics approval prior to data collection. 

\subsection{System Implementation}
\label{sec:implementation}




MindFort was implemented using the Flask framework for the Web, containerised for reproducibility, and hosted locally on a secure private institutional server as a public web application.
MindFort's Conversational Inoculation (Forty the chatbot) was based on the gpt-4o-2024-11-20\footnote{\url{https://openai.com/index/hello-gpt-4o/}} model with the temperature parameter set to 1. 

\begin{figure}[htpb]
    \centering
    \frame{
        \includegraphics[width=0.48\textwidth]{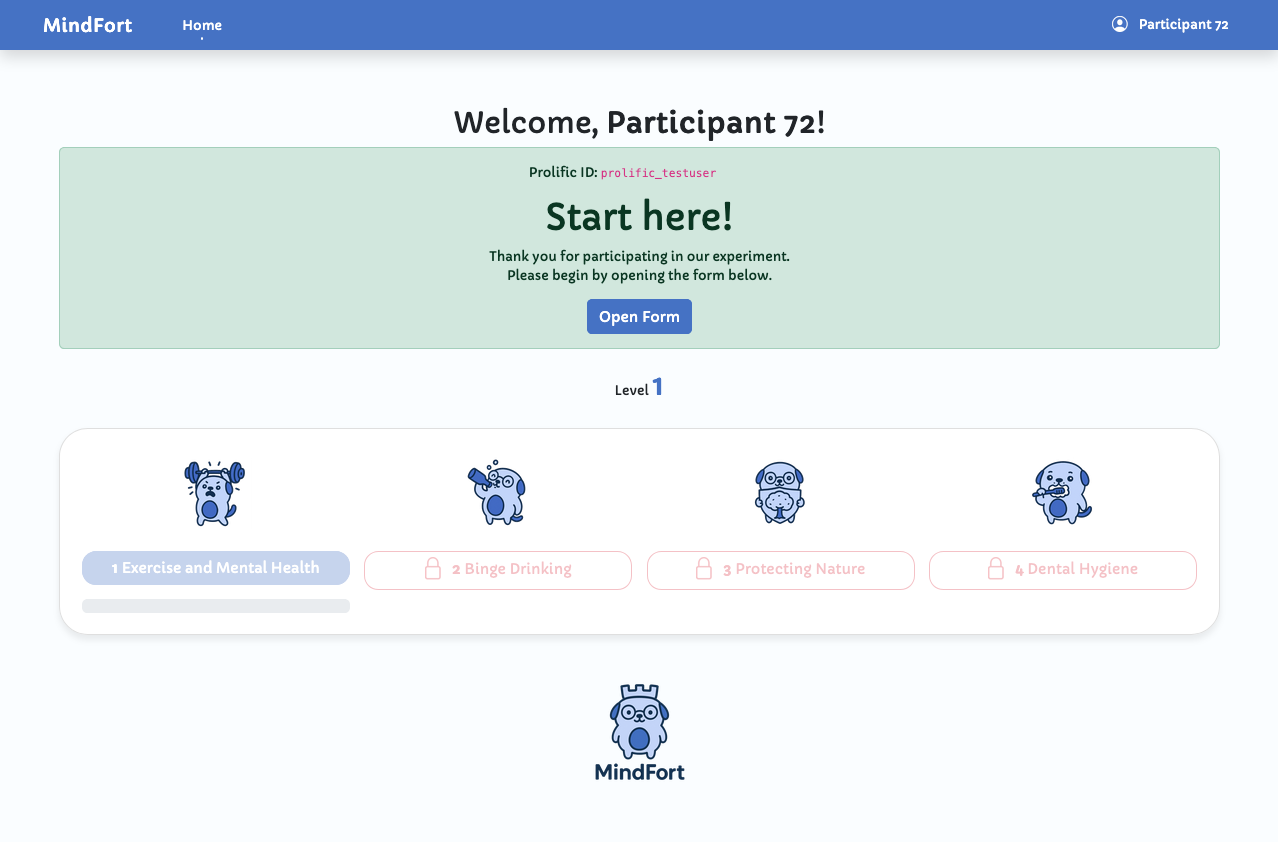}
    }
    \caption{Prototype Screenshot (1/7): Home page of the signed-in user showing their lesson progress. In this image, all lessons are still locked as the participant has not opened the initial form yet by clicking the ``Open Form'' button.}
    \Description{Screenshot of the MindFort system's home page, showing the user's lesson progress. The page welcomes "Participant 72." In the center is a "Start here!" section with an "Open Form" button. Below this four lesson modules are displayed horizontally. The first lesson, "1 Exercise and Mental Health," is highlighted as the current, active lesson. The other three lessons, "2 Binge Drinking," "3 Protecting Nature," and "4 Dental Hygiene," are grayed out and have a padlock icon, indicating they are locked.}
    \label{fig:ss1}
\end{figure}

The front page first redirected participants to an external form, where participants provided consent. 
The page layout is depicted in \autoref{fig:ss1}.
Short titles of the topics for the four \textit{lessons} were listed, where only the first lesson was accessible at first.
The remaining three were displayed as “locked,” indicated by a lock symbol on their start buttons.
The users always started the experiment with the first lesson.



\subsubsection{Lessons}

In MindFort, each Lesson serves as a unit of inoculation, targeting a single misinformation topic.
The four topics and the related misconceptions are summarised in Appendix C.
Each of the four lessons consisted of five distinct \textit{stages}, which together implement an inoculation study, as follows..

\subsubsection{Stages}
\label{sec:implementation_stages}

The stages in each leasson implement an inoculation study.
An inoculation study targeting misinformation typically proceeds in a series of well-defined steps \cite{McGuire_1964, Fransen_Mollen_Rains_Das_Vermeulen_2024, godbold2000}.
First, participants report their certainty about a truthful claim, i.e. a factually correct statement. 
This step is important to ensure that participants agree with the truthful claim at the beginning. 
Second, they engage in a \textit{treatment}, where in some way the participants learn about aspects of, or claims about, the topic that are commonly misunderstood or deliberately misrepresented. 
Participants can, for example, research and argue against a ``dose of weakened misinformation'', strengthening their resistance to stronger forms of persuasion in the future. 
Or they may receive education about the topic, called the supportive defense treatment.
Other forms can exist too, and naturally, this treatment can be realised in many ways.
The purpose of treatment is to better equip people to resist a piece of misinformation in the future.
As the third step, participants again report their certainty about the same original truthful claim, thus providing a measure of the effect of the treatment itself, which also contains misinformation in one way or another (study, argue against, etc.).
Fourth, participants encounter a strong counterattitudinal attack arguing against their currently held truthful beliefs, typically in the form of a persuasive piece of misinformation (typically custom-written or drawn from online sources in case of textual misinformation).
In the final step, participants report their certainty about the original claim for the third and final time. 
Comparing participants' certainty scores across these three points (pre-treatment, mid-lesson, and post-attack) reveals how much the inoculation method helped participants to build resistance against the stronger misinformation attack.

\begin{figure}[htpb]
    \centering
    \frame{
        \includegraphics[width=0.4\textwidth]{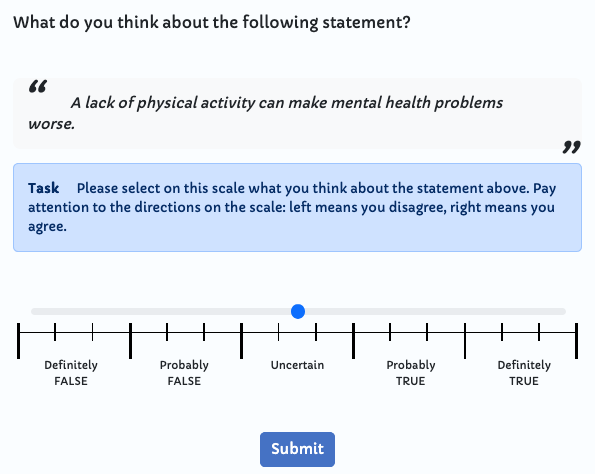}
    }
    \caption{Prototype Screenshot (2/7): Stages 1, 3 and 5 involve a 15-point scale measuring the participant's certainty concerning a truthful claim.}
    \Description{Partial screenshot of the certainty rating interface. The interface asks the user, "What do you think about the following statement?" followed by the statement itself: "A lack of physical activity can make mental health problems worse." A blue task box provides instructions: "Please select on this scale what you think about the statement above. Pay attention to the directions on the scale: left means you disagree, right means you agree." Below this is a 15-point horizontal scale with five main labels. From left to right, these are: "Definitely FALSE," "Probably FALSE," "Uncertain," "Probably TRUE," and "Definitely TRUE." A "Submit" button is centered below the scale.}
    \label{fig:ss4}
\end{figure}

In our study, the 5-step inoculation process described above was implemented in five Stages:
\begin{enumerate}
    \item \textbf{Pre-treatment Certainty Score} presents users with a truthful claim and asks them to rate their certainty about it on a 15-point scale that is identical to that of \citet{McGuire_1961} (see \autoref{fig:ss4}).
    \item \textbf{Treatment} displays one of the three treatments our study implemented, i.e. \textbf{Reading}, \textbf{Writing}, and conversational inoculation: \textbf{Chatbot}. Thus, this stage serves to the user one of the conditions, including a \textbf{Control} condition in which this step is skipped entirely (no inoculation).
    \item \textbf{Mid-Lesson Certainty Score} repeats the certainty score question, re-displaying the original truthful claim for the user to consider again.
    \item \textbf{Strong Counterattitudinal Attack} presents a five-paragraph, convincingly written text (see Appendix J) arguing \textit{against} (attacking) the truthful claim presented earlier and the participant's own opinion, which are presumed to align. This step simulates the effect of a believable misinformation piece encountered online with a professional writing style and added misused citations to back up incorrect claims. 
    \item \textbf{Post-Attack Certainty Score and IMI questionnaire} again repeats the certainty score question, with an additional Internal Motivation Inventory (IMI) questionnaire including questions for four subscales: \textit{Interest/Enjoyment}, \textit{Perceived Competence}, \textit{Effort/Importance} and \textit{Value/Usefulness}.
\end{enumerate}

Due to the nature of crowdwork tasks, we implemented no delay between inoculation and the strong counterattitudinal attack, although the increased resistance is typically more pronounced after a delay \cite{godbold2000}. 


\subsubsection{Study conditions}
\label{sec:conditions}
MindFort implemented three distinct treatments (study conditions): \textit{Reading}, \textit{Writing}, and \textit{Chatbot}.
The Reading condition implements a simple supportive defense treatment (i.e. only reading, and only supportive materials, no misinformation in any form). Writing implements a traditional refutation method for cognitive inoculation by independent research and writing. The Chatbot condition implements the Conversational Inoculation approach as a dialogue between the participant and a chatbot as a proactive debate partner, which that was the focus of our investigation.
Additionally, we added a control condition to collect data about participants' baseline resistance to misinformation. 

%

\textbf{The Reading condition} involved reading a five-paragraph essay (see Appendix D) arguing \textit{for} the truthful belief related to the current lesson topic. Study participants were required to read the essay for at least three minutes. 
The essays were structured in line following \citeauthor{Fransen_Mollen_Rains_Das_Vermeulen_2024}'s example \cite{Fransen_Mollen_Rains_Das_Vermeulen_2024}. 
The first paragraph briefly introduces the topic, then the following four paragraphs support four correct beliefs (corresponding to the four misconceptions in the other conditions and in the strong counterattitudinal attack) regarding the subject, using external resources, but it does not directly address misconceptions or refute misinformation. 
All four of the essays were prepared by the first author and reviewed by one co-author, and each of them always addressed the same four misconceptions that the other conditions and the counterattitudinal attack for the same lesson did.

\textbf{The Writing condition} presented the participants with a list of four misconceptions and an input field to write a short essay refuting each of them (see \autoref{fig:herocollage}/a or Appendix A, Fig. 11).
The task specifically asked participants to use external resources but to avoid using AI.
We also posed a lower limit of 400 characters but no time limit for the essay writing task. 
Again, this simple task is in line with traditional Cognitive Inoculation experiments \cite{McGuire_1961, Fransen_Mollen_Rains_Das_Vermeulen_2024}.

\textbf{The Chatbot condition} (see \autoref{fig:herocollage}/b or Appendix A, Fig. 10.) instantiates Conversational Inoculation. 
It displays a chat window familiar from online chatbots or mobile chat applications.
The user is asked to initiate the conversation, but the chatbot is designed to be in control of the lesson and proactively lead the refutation-building progress. 
The chatbot embeds threat and refutational preemption \cite{Banas_Rains_2010} in the conversation (see Appendix B). It does that by appearing to take a counter-attitudinal position at first and by proactively guiding the participants to disprove the misconceptions shared by the chatbot. This is exemplified with the conversations of Participants 9 and 57 in Appendix H. As opposed to regular conversations with LLM-based chatbot assistants such as Claude or Qwen Chat, the chatbot has a strong lead on the conversation. It also strongly adheres to the goals Conversational Inoculation, even when facing resistance such as the user disengaging or attempting to derail the discussion.
To this end, the system prompt of the chatbot in the prototype (called Forty) was engineered in line with literature on prompting and an inoculator persona. 
The complete system prompt along with annotations referring to the literature informing the prompt design is included in Appendix B. 
Based on \citeauthor{White_Fu_Hays_Sandborn_Olea_Gilbert_Elnashar_SpencerSmith_Schmidt_2023}'s work \cite{White_Fu_Hays_Sandborn_Olea_Gilbert_Elnashar_SpencerSmith_Schmidt_2023}, we defined a Persona and prepared the bot for a flipped interaction \cite{chen2025experience} as the chatbot is expected to lead the conversation, proactively challenging the participant's truthful beliefs and help them develop their defences. Further, based on \citeauthor{Giray_2023}'s recent article \cite{Giray_2023} which gathered useful practical advice for prompt engineering, we paid particular attention the amount of task context disclosed to the language model, how we deal with the LLMs' inherent limitations, and the level of clarity in the prompt phrasing, as shown in Appendix B. \new{As one of the core benefits of the proposed inoculation method, we aimed to leverage the flexibility LLMs offer by supported the adaptability of the agent within the reasonable boundaries of the task context. This was realised with the system prompt in multiple ways. First, we placed the agent in the role of a partner rather than an educator. Second, it was given instructions to build the conversation around the participant's own reasoning to reinforce those views rather than searching for new angles on the topic. Third, the agent was prompted with a goal-oriented approach rather than giving the agent explicit instructions on how to induce the desired effect.}

\textbf{The Control condition} did not implement any treatment, i.e. the participant simply skipped the stage and proceeded directly to the Mid-lesson Certainty Score stage.

\subsection{Experimental Design}

We conducted a within-subjects experiment online.
The within-subject design was chosen because it mitigates the large individual differences in susceptibility to persuasion, a critical confounder in this research area.
It allows for a direct, within-participant comparison of the user experience across the different inoculation methods, and it makes more efficient use of the participant pool given the significant briefing and debriefing overhead.
\old{The experiment was designed to evaluate the effectiveness of Conversational Inoculation compared to other treatment approaches (RQ1), to measure participants’ intrinsic motivation to engage with Conversational Inoculation (RQ2), and to explore additional factors that might promote or hinder its effectiveness (RQ3).}
\new{The experiment was designed to evaluate the effectiveness of Conversational Inoculation compared to other treatment approaches (RQ1), to measure participants’ intrinsic motivation to engage with Conversational Inoculation (RQ2), and to explore additional factors or cognitive processes that contribute to and make up the Conversational Inoculation process (RQ3).}

Concerning RQ1, the independent variable was the type of treatment administered, and the dependent variable was the change in participants’ certainty scores given on a 15-point scale (see \autoref{fig:studydesign}).
For RQ2, we assessed intrinsic motivation of the users using the Intrinsic Motivation Inventory (IMI) \cite{ryan1982control}, with subscales for \textit{Interest/Enjoyment}, \textit{Perceived Competence}, \textit{Effort/Importance}, and \textit{Value/Usefulness}. 
These subscales allow us to understand to what extent participants were motivated to participate in Conversational Inoculation.
Simultaneously, these metrics may inform future development needs in the MindFort system itself, and enable comparison between the study conditions.
Finally, to answer RQ3, we logged all participants’ chats to be analysed quantitatively and qualitatively to explore linguistic features and mechanisms that might promote or inhibit Conversational Inoculation.
To that end, we conducted a Linguistic Inquiry and Word Count (LIWC) analysis of the chatbot conversations using the LIWC-22 (version 1.11.0) analysis tool \cite{Pennebaker2022}.
More specifics are provided in the data analysis section later.

To ground our study in areas where misinformation is prevalent and has a significant impact, we focused on the broad theme of wellbeing. We selected three health–related topics on which most participants could be expected to hold accurate prior beliefs: (1) \textit{Exercise and Mental Health}, (2) \textit{Binge Drinking}, (3) \textit{Dental Hygiene}. To explore the generalisability of Conversational Inoculation, we also included a fourth topic: (4) \textit{Protecting Nature}.
We collected certainty scores, IMI questionnaire responses, open feedback, and logs of participants' chats and written essays across all these topics.

\new{To maintain a reasonable session duration given this extensive data collection, we did not directly measure cognitive load via physiological metrics or questionnaires (e.g., NASA-TLX). Instead, we mitigated potential fatigue through our experimental setup. Counterbalanced condition ordering distributed fatigue effects evenly, and we kept individual interaction tasks short, approximately 5--10 minutes.}

These design choices were validated in a pilot experiment with four participants. The pilot helped identify browser support issues, task timing problems, flaws in the bot conversations, and other miscellaneous aspects that were improved in the final version of the experiment and the prototype itself.

\subsection{Participants}
We recruited participants from \textit{Prolific}, a human subject pool offering comprehensive pre-screening filters and high-quality responses from trusted participants~\cite{Peer2017BeyondResearch, Douglas2023DataSONA}.
Each participant was compensated £12, which, at the average 75-minute completion time, is £10.70 per hour. Four participants who worked significantly overtime received bonus payments of £2 to £5 afterwards.
Participants were required to be from predominantly English-speaking countries (UK, Ireland, US, Australia, NZ), have submitted at least 200 tasks previously, have an at least 90\% task approval rate, use a desktop computer for the experiment, and use English as their first language.

We determined a minimum sample size of 57 participants using G*Power 3.1 for the Wilcoxon signed-rank test, based on a small effect size ($d = .5$) and an alpha level of 0.05. A larger sample size ($N = 65$) was used in our study to account for potential participant exclusion and ensure adequate statistical power for our analysis. \new{The study experienced 9 dropouts. Two participants dropped out when shown the Writing task, one when shown the Chatbot task, and six before beginning their first task.}

\subsection{Procedure}
\label{sec:procedure}

\begin{figure}[htpb]
    \centering
    \includegraphics[width=0.5\textwidth]{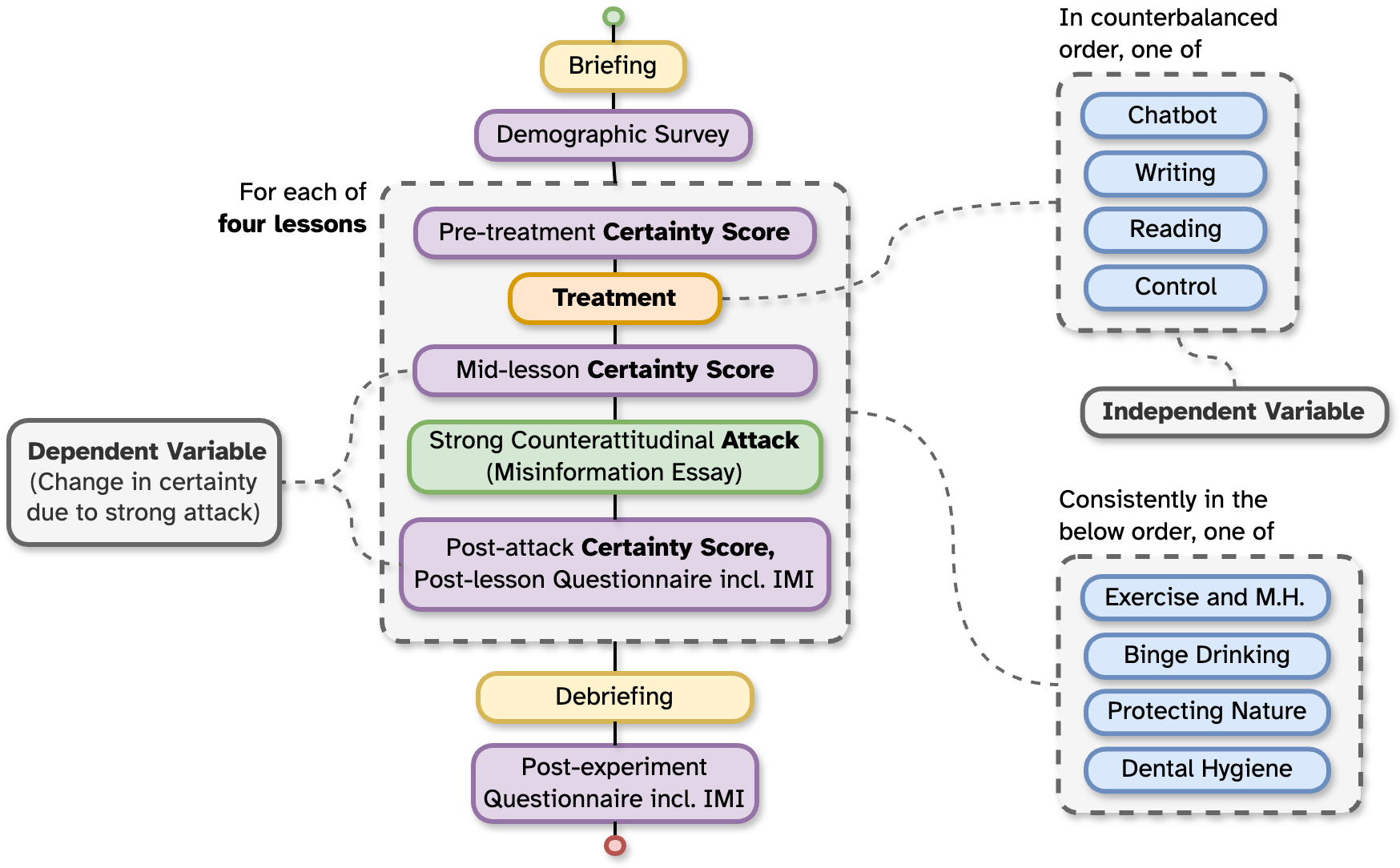}
    \caption{Diagram illustrating our experiment with additional annotations. Each participant completes four lessons consisting of five stages each. The second stage represents the Independent Variable, one of four possible conditions. The change in the participant's self-reported certainty in their truthful belief before and being exposed to misinformation is our Dependent Variable. After all four lessons are completed, the participants get debriefed and informed about the purpose of the study first and then they proceed to fill out Post-Debriefing IMI questionnaire.}
    \Description{Flowchart illustrating the experimental procedure. The main flow starts with a "Briefing" and a "Demographic Survey." This is followed by a process that is repeated for each of four lessons. The per-lesson process consists of five steps in sequence: Pre-treatment Certainty Score,Treatment, Mid-lesson Certainty Score, Strong Counterattitudinal Attack (Misinformation Essay), Post-attack Certainty Score. An annotation explains the "Independent Variable." The "Treatment" step is linked to a box showing the four possible conditions, which are applied in a counterbalanced order: "Chatbot," "Writing," "Reading," and "Control." Another annotation explains the "Dependent Variable," which is the "Change in certainty due to strong attack." This is calculated from the difference between the "Mid-lesson Certainty Score" and the "Post-attack Certainty Score." A third annotation shows the four lesson topics, which are presented in a consistent order: "Exercise and Mental Health," "Binge Drinking," "Protecting Nature," and "Dental Hygiene." After all four lessons are completed, the experiment concludes with a "Debriefing" and a final "Post-experiment Questionnaire including IMI."}
    \label{fig:studydesign}
\end{figure}

The procedure is visualised in \autoref{fig:studydesign}.
Participants, upon clicking the study URL in Prolific, were shown the home page of an automatically registered and signed-in user based on their unique Prolific user ID (see \autoref{fig:ss1}). 

Each participant then completed four lessons.
Every lesson was themed around one of the four topics.
In a lesson, participants completed the inoculation process following one of the four methods. 
The four conditions were assigned in a counterbalanced Latin Square order to mitigate carryover effects.
The topic ordering was consistent (i.e. Lessons 1,2,3,4 were around the same topic for all participants).
After participants gave their informed consent, they proceeded to complete the four lessons, in the five stages explained earlier in \autoref{sec:implementation_stages}.
Thus, each participant completed all four lessons with three treatment approaches and Control.

After the fourth lesson, using a task completion code (Appendix A, Fig. 14.), participants could use to proceed into the debriefing (see Appendix F) that is designed to mitigate any confusion and harm caused by the misinformation shown to participants. 
This debriefing text was also shown to participants who decided not to finish the the experiment. 
After debrefing, the participants filled out the IMI form again. 
They then had a final chance to share any thoughts or feedback, and with that, the experiment was concluded.

\subsection{Data Analysis Methods}

For the quantitative analysis, we followed standard statistical methods in HCI, first testing for data normality and then selecting appropriate subsequent methods, applying pairwise corrections when necessary.
The Results section (\autoref{sec:results}) provides additional details on the analysis.

For analysing the chats and open-ended items, we employed Conventional Content Analysis~\cite{hsieh_three_2005}.
We chose a deductive approach for our qualitative analysis. This was because the scope for answering RQ2 and RQ3 was defined in advance, focusing on user experiences with Conversational Inoculation and the factors that may promote or inhibit it.
The first author of the article initially coded a subset of the data.
This code set was then refined in collaboration with a co-author to ensure clarity and comprehensiveness.
The first author then systematically applied these codes to all messages in the dataset.
This message-level coding allowed us to identify patterns in how codes co-occurred and sequenced throughout a conversation.
The final analysis focused on interpreting these patterns to develop overarching themes.

\section{Results}
\label{sec:results}
\subsection{Participant Demographics}

We recruited 65 participants, excluding four who failed an attention check, resulting in a final sample of 61 participants. The participants' ages ranged from 19 to 73 (M = 37.46, SD = 14.21), and 55.7\% identified as women. The group was predominantly White (65.6\%) and lives in the United States (62.3\%) and the United Kingdom (29.5\%). Most participants (63.9\%) were not students. A comprehensive breakdown of all demographic data, including ethnicity, nationality, student status, and employment status, is presented in Appendix G.



\subsection{Effectiveness of Conversational Inoculation (RQ1)}

To investigate whether the Conversational Inoculation treatment implemented in MindFort reduced susceptibility to counterattitudinal attacks, we first focused on comparing the post-attack certainty changes between the Chatbot condition and the other experimental conditions.
Further, in inoculation research, individual susceptibility to persuasion is a key confounder. 
We 
accounted for this by using the Control condition as a within-subject baseline, capturing each participant's natural susceptibility to misinformation in the absence of inoculation.
We note that these analyses address two different underlying aspects of RQ1.
The former addresses whether treatments differ in group-level differences in effectiveness.
The latter examines the unique contribution of each inoculation method, beyond a participant's natural resistance.
Prior to analysis we removed outliers using the $\pm$1.5 × IQR threshold \cite{tukey1977exploratory}.

\subsubsection{Post-attack certainty change}
\label{sec:rq1_apriori}
The participants entered how certain they were about the subject's truthful claim using a 15-point certainty scale three times for each lesson. 
Here, the attack refers to the \textit{Strong Counterattitudinal Attack}, which presented a misinformation essay to the participants (see \autoref{fig:studydesign}).
The post-attack certainty change is thus calculated as the \textit{difference between the mid-lesson certainty score and the post-attack certainty score}:

\begin{equation}
\label{eq:postattackcertaintychange}
\text{Post-attack Certainty Change} = \text{certainty}_{post} - \text{certainty}_{mid}
\end{equation}

The certainty scores were collected on a 15-point scale and are therefore treated as ordinal data.
Consequently, we used non-parametric statistical methods.
A Friedman rank sum test did not indicate significant differences between conditions ($\chi^2(3) =6.5$, $p = .09$). 
Because our focus was on contrasts involving the Chatbot condition, and omnibus tests may fail to account for such differences when other conditions are highly similar, we next conducted Bonferroni-corrected pairwise Wilcoxon signed-rank tests comparing Chatbot with each of the other conditions.



\begin{figure*}[tb]
    \centering
    \begin{minipage}[t]{0.48\linewidth}
        \centering
        \includegraphics[height=4cm]{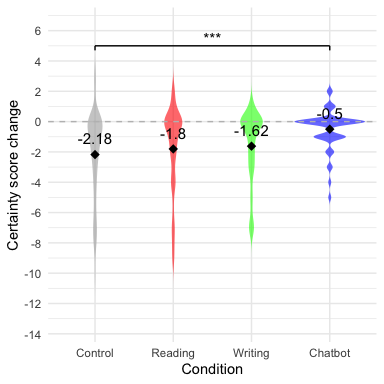}
        \caption{Post-attack certainty change compared between the four conditions. The amount of certainty change shows how much participants' certainty increased or decreased on the 15-point certainty scale after exposing them to misinformation. A lower certainty change score means higher resistance to misinformation.}
        \Description{A violin plot comparing the "Certainty score change" on the y-axis for four experimental conditions on the x-axis. The conditions and their mean certainty changes are: Control (-2.18), Reading (-1.8), Writing (-1.62), and Chatbot (-0.5). A bracket with three asterisks indicates a statistically significant difference between the Control and Chatbot conditions.}
        \label{fig:inoc_outcomes}
    \end{minipage}
    \hfill
    \begin{minipage}[t]{0.48\linewidth}
        \centering
        \includegraphics[height=4cm]{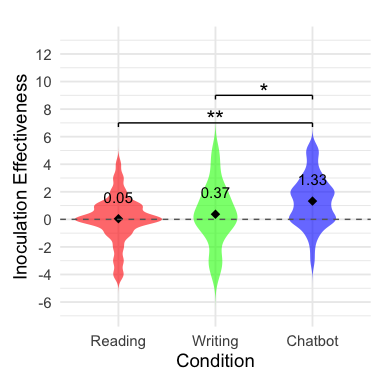}
        \caption{Method-wise Inoculation Effectiveness, i.e. Certainty score change when controlling for individual susceptibility, as defined in \autoref{eq:Inoculation_Effectiveness}). Higher values mean greater impact on one's baseline resistance to misinformation.}
        \Description{A violin plot comparing the "Inoculation Effectiveness" on the y-axis for three experimental conditions on the x-axis. The conditions and their mean effectiveness scores are: Reading (0.05), Writing (0.37), and Chatbot (1.33). Brackets with asterisks indicate that the Chatbot condition is significantly more effective than both the Reading condition (two asterisks) and the Writing condition (one asterisk).}
        \label{fig:effectiveness}
    \end{minipage}
\end{figure*}

The pairwise comparisons revealed a statistically significant difference between the Control condition and the Chatbot condition ($p = .001$, $r = -.33$), as depicted in \autoref{fig:inoc_outcomes}.
Participants in the Chatbot condition ($M = -0.5$, $SD = 1.32$) showed less certainty change compared to Control ($M= -2.18$, $SD = 2.94$), indicating substantially greater resistance to the misinformation attack.
Pairwise comparisons between the Chatbot and Reading ($M = -1.8$, $SD = 2.89$) conditions ($p = .07$, $r = .2$) and the Chatbot and Writing ($M = -1.62$, $SD = 2.43$) conditions ($p = .08$, $r = .2$) showed no significant difference. 
To further investigate RQ1, we next controlled for individual differences in overall susceptibility. 

\subsubsection{Controlling for individual differences in susceptibility}
\label{sec:rq1controlresults}

We used the participants' Post-attack certainty score changes in the Control condition as a baseline resistance to misinformation to account for individual differences in susceptibility among participants. 
By doing so, we can compare how much each treatment affected subjects' resistance to persuasion compared to their baseline.
We define this as Inoculation Effectiveness:

\begin{equation}
\label{eq:Inoculation_Effectiveness}
\begin{gathered}
\text{Inoculation Effectiveness} = \\
(\text{certainty}^{control}_{post} - \text{certainty}^{control}_{mid}) \\
- (\text{certainty}^{treatment}_{post} - \text{certainty}^{treatment}_{mid})
\end{gathered}
\end{equation}

Using a Shapiro-Wilk normality test, we determined that Writing and Chatbot conditions do not significantly deviate from normal distribution, however the Reading condition does.
Therefore, we used non-parametric tests. 

A Friedman rank sum test did not show a significant effect ($\chi^2(2) = 0.89$, $p = .64$).
Following our analysis plan for RQ1, we then proceeded with Wilcoxon pairwise tests (Bonferonni corrected), which showed significant differences.
The Chatbot condition showed higher Inoculation Effectiveness compared to both the Reading condition ($p = .004$) and the Writing condition ($p = .033$), as depicted in \autoref{fig:effectiveness}.

\subsubsection{Overall certainty change due to the treatment}

We compared how the different treatments affected participant's certainty changes. 
This was done by comparing the \textit{difference of mid-lesson and pre-treatment certainty scores} for each condition.
Though treatment (threat) is a part of a successful inoculation treatment, with too much threat (i.e., too strong a misinformation message during treatment), we risk negatively influencing the subject's certainty.
The mean certainty score changes were: Control $=0.00$, Reading $=0.29$, Writing $=-0.34$, and Chatbot $=0.24$. 
A Friedman rank sum test did not show significant differences between the conditions ($\chi^2(3)=6$, $p=.1$).
Wilcoxon rank sum tests further verified the difference between the Chatbot and Writing conditions ($r=-.30$, $p=.006$). Notably, in our case, Conversational Inoculation with Forty the chatbot \textit{increased} participant's certainty, while the treatment in the Writing condition \textit{decreased} it.

We also compared the \textit{difference of post-attack and pre-treatment certainty scores} for each condition to understand how our experiment as a whole affected participants' certainty scores, as negative impact on participants was an ethics concern.
The mean certainty score changes were: Control $=-2.23$, Reading $=-0.54$, Writing $=-1.64$, and Chatbot $=-1.35$.
A Friedman rank sum test detected significant differences between the conditions ($\chi^2(3)=11$, $p=.01$), but Wilcoxon rank sum tests did not find significant pairwise differences between conditions after Bonferroni correction.

The negative certainty changes across all conditions suggest that our counterattitudinal messages were highly effective. 
This was an anticipated outcome and was the reason for the inclusion of the Debriefing stage, where participants were explained the study specifics and reassured about their initial beliefs. 


\subsubsection{Topic-level analysis}
\label{sec:topic_differences}

\begin{figure*}[tb]
    \centering
    \begin{minipage}[t]{0.48\linewidth}
    \centering
    \includegraphics[width=0.9\linewidth]{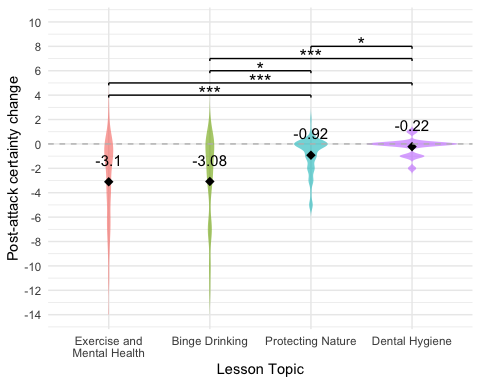}
    \caption{Post-attack certainty change compared between the four topics. The amount of certainty change shows how much participants' certainty increased or decreased on the 15-point certainty scale after exposing them to misinformation. A lower certainty change score means higher resistance to misinformation.}
    \Description{A violin plot comparing the "Certainty score change" on the y-axis across the four different lesson topics on the x-axis. The topics and their mean certainty score changes are: Exercise and Mental Health (-3.1), Binge Drinking (-3.08), Protecting Nature (-0.92), and Dental Hygiene (-0.22). Brackets with asterisks indicate five statistically significant differences between the topics. Both "Exercise and Mental Health" and "Binge Drinking" show significantly larger decreases in certainty compared to both "Protecting Nature" and "Dental Hygiene." Additionally, "Protecting Nature" shows a significantly larger decrease than "Dental Hygiene."}
    \label{fig:acrosstopics}
    \end{minipage}
    \hfill
    \begin{minipage}[t]{0.48\linewidth}
        \centering
    \includegraphics[width=0.9\linewidth]{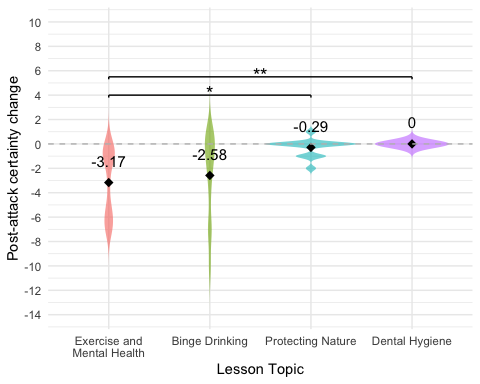}
    \caption{Post-attack certainty change compared between the four topics in the Chatbot condition.}
    \Description{A violin plot comparing the "Certainty score change" on the y-axis across the four different lesson topics on the x-axis. The topics and their mean certainty score changes are: Exercise and Mental Health (-3.17), Binge Drinking (-2.58), Protecting Nature (-0.29), and Dental Hygiene (0). Brackets with asterisks indicate five statistically significant differences between the topics. Both "Exercise and Mental Health" and "Binge Drinking" show significantly larger decreases in certainty compared to both "Protecting Nature" and "Dental Hygiene." Additionally, "Protecting Nature" shows a significantly larger decrease than "Dental Hygiene."}
    \label{fig:acrosstopics_chatbot}
    \end{minipage}
\end{figure*}

All participants engaged with four distinct topics during the study.
Three related to health and one related to nature.
In the following, we analyse differences in the inoculation outcomes between topics. An overview of the Certainty change after the Strong Counter-attitudinal Attack can be observed in \autoref{fig:acrosstopics}.

As previously established, post-attack certainty score changes and inoculation effectiveness were not normally distributed.
Here, a Friedman test showed a statistically significant difference between post-attack certainty change across topics ($\chi^2(2) = 21$, $p < .001$).
Post-hoc Wilcoxon tests further revealed that the post-attack certainty change was different in several cases. 
Exercise and Mental Health showed more decrease in certainty than both Protecting Nature ($p = .001$, $r = -.34$) and Dental Hygiene ($p < .001$, $r = -.49$).
Likewise, there was a more substantial post-attack certainty change in the Binge Drinking topic than in both Protecting Nature ($p = .02$, $r = -.29$) and Dental Hygiene ($p < .001$, $r = -0.45$) topics. 
Protecting Nature showed slightly more certainty decrease than Dental Hygiene ($p = .039$, $r = -.25$). 
No difference was observed between Exercise and Mental Health and Binge Drinking ($p = 1.0$, $r = -.02$). 

When focusing on the Chatbot condition alone (\autoref{fig:acrosstopics_chatbot}, certainty decreased more strongly in the Exercise and Mental Health lesson compared to both Protecting Nature ($p = .013$, $r = -.54$) and Dental Hygiene ($p = .002$, $r = -.64$).

To account for variation across both participants and topics, we fitted a linear mixed-effects model using the \texttt{lme4} package in R \cite{lme4}, with condition as a fixed effect and participants and topics as random effects.
In the Control condition, certainty scores decreased by 1.72 points on average ($SE = 0.48$, $t=-3.57$, $p=.009$), however the treatment conditions were not found to be significantly different from Control (see \autoref{tab:lmem}), which can be explained by the significant variance due to both participant differences ($\sigma^2 = 1.50$) and topic differences ($\sigma^2 = 0.54$).
These findings further validate that the inoculation effect was not uniform across the topics we studied.

\begin{table}[htpb]
\caption{Fixed Effects from the Linear Mixed-Effects Model for Post-attack certainty change.}
\label{tab:lmem}
\centering
\small
\begin{tabular}{lrrrr}
\toprule
\textbf{Fixed Effect} & \textbf{Estimate} & \textbf{Std. Error} & \textbf{\textit{t}} & \textbf{\textit{p}} \\
\midrule
Intercept [Control] & -1.72 & 0.483 & -3.57 & .009 \\
Reading & 0.45 & 0.39 & 1.17 & .242 \\
Writing & 0.04 & 0.39 & 0.10 & .922 \\
Chatbot & 0.32 & 0.38 & 0.86 & .393 \\
\bottomrule
\end{tabular}
\end{table}


\subsection{Intrinsic Motivation (RQ2)}

\begin{figure*}[tb]
    \centering
    \includegraphics[width=0.9\linewidth]{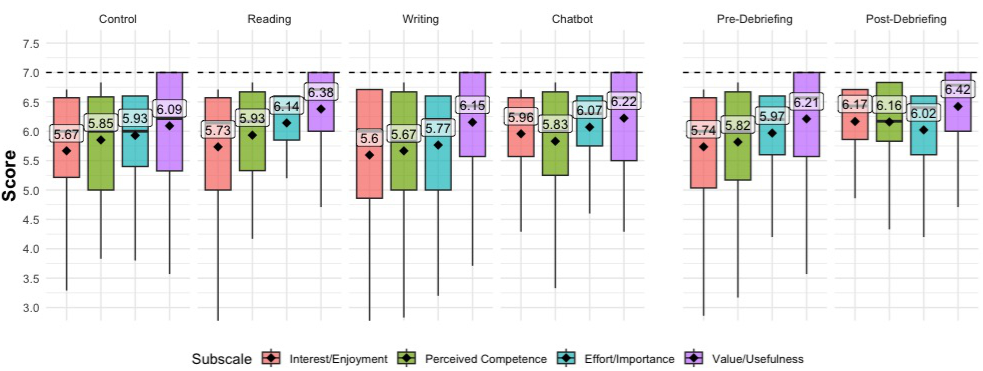}
    \caption{IMI scores for each of the four conditions (a--d), Pre-Debriefing scores calculated as the overall IMI scores from all four conditions (f) and Post-Debriefing IMI scores (g) for each of the four measured subscales.}
    \Description{A series of box plots showing the IMI scores on the y-axis for six different groups on the x-axis. Each group contains four box plots, corresponding to the four IMI subscales: Interest/Enjoyment, Perceived Competence, Effort/Importance, and Value/Usefulness. The mean scores for each condition are as follows: Control Condition: Interest/Enjoyment is 5.67, Perceived Competence is 5.85, Effort/Importance is 5.93, and Value/Usefulness is 6.09. Reading Condition: Interest/Enjoyment is 5.73, Perceived Competence is 5.93, Effort/Importance is 6.14, and Value/Usefulness is 6.38. Writing Condition: Interest/Enjoyment is 5.6, Perceived Competence is 5.67, Effort/Importance is 5.77, and Value/Usefulness is 6.15.    Chatbot Condition: Interest/Enjoyment is 5.96, Perceived Competence is 5.83, Effort/Importance is 6.07, and Value/Usefulness is 6.22. Pre-Debriefing Scores: Interest/Enjoyment is 5.74, Perceived Competence is 5.82, Effort/Importance is 5.97, and Value/Usefulness is 6.21. Post-Debriefing Scores: Interest/Enjoyment is 6.17, Perceived Competence is 6.16, Effort/Importance is 6.02, and Value/Usefulness is 6.42.}
    \label{fig:imi-subscales}
\end{figure*}

IMI scores were collected at the end of each lesson, which encompassed the full five-step lesson procedure rather than the treatment alone.
Thus, the measures reflect participant experience of the complete lesson around a given topic and treatment method. \autoref{fig:imi-subscales}
 depicts the four collected subscales for the four conditions, their averages and the post-debriefing IMI scores. 

Using a Shapiro-Wilk normality test, we determined that none of the IMI subscales follow normal distribution ($p < .001$).

No statistically significant differences were observed between conditions on any subscale (\textit{all p} $>$ 0.05), indicating that
all treatments were perceived as approximately equally valuable and engaging.

A Friedman test indicated differences among conditions on the Perceived Competence subscale ($\chi^2(3) = 9.40$, $p = .02$), but pairwise Wilcoxon comparisons did not reveal significant differences between any conditions.

We also compared the mean IMI scores of the four conditions to the Post-Debriefing IMI scores to see how participants' view of the system changed when they were able to form a potentially more informed opinion.
Wilcoxon signed-rank test showed a significant increase in Interest/Enjoyment scores ($p < .001$), Perceived Competence ($p< .001$), and Value/Usefulness ($p<.001$) but not in Effort/Importance ($p = .10$).

\subsubsection*{Open feedback}
\label{sec:feedback}

All responses were reviewed to identify positive and negative aspects of the user experience. Most left feedback at least once, but typically no more than two times out of the 5 opportunities.

Participants praised the chatbot, with one noting it \textit{``feels more interactive and it encourages me to learn faster''} (P51, Chatbot condition), and P15 (Post-debriefing) calling the conversation with \textit{``Forty''} the \textit{``most interesting and significant aspect of the entire system''}. This positive experience was often linked to a sense of \textit{``gamified''} learning that was \textit{``entertaining AND interesting''} (P65, Chatbot condition), which helped the study's core goal of promoting critical reflection. Participants across all treatment conditions reported that the tasks successfully challenged their preconceptions and led to the acquisition of new knowledge, with P26 and P47 (Chatbot condition), and P63 (Post-debriefing) all stating the system \textit{``made me think more deeply''} The system was described as easy to navigate and seamless (P63, Post-debriefing and P60, Chatbot condition). 

The above positive sentiment was contrasted by the non-chatbot conditions. The static nature of the control condition was found to be \textit{``very boring''} (P27, Control condition), while the writing task was described as \textit{``more exhaustive''} (P65, Writing condition) and stressful, with some users feeling they were \textit{``really pressed for time''} (P37, Writing condition) to conduct research and formulate refutations. Participants experienced interactional friction, such as a bug that caused the chat window to scroll to the top after each message, timer glitches and, for some, the chatbot's perceived unnaturalness. Even when functioning correctly, its behavior sometimes broke the sense of a genuine conversation, as users found it \textit{``unnatural and a bit awkward to use''} (P60, Post-debriefing) or felt it \textit{``wanted to end the conversation sooner than I would have liked''} (P49, Post-debriefing). Participant 21 (Chatbot condition) found the chatbot intellectually unconvincing, noting it \textit{``wasn't the cleverest I've ever spoken to''} and that it \textit{``seemed to change its position''} , undermining the credibility of the challenge. Participant 65 (Post-debriefing) reported that they \textit{``did lose a little confidence in my belief''}, highlighting that the inoculation process itself can be an uncomfortable and confidence-shaking experience.

\subsection{\old{Factors Promoting and Inhibiting Inoculation (RQ3)} \new{Factors Affecting Inoculation (RQ3)}}


\subsubsection{Qualitative analysis of chatbot conversations}
\label{sec:qualchatbotconv}

A qualitative analysis of the chatbot conversations helps understand key factors 
that promote or inhibit Conversational inoculation.

\paragraph{Adapting the inoculation approach}
We designed the chatbot to be goal-oriented but set very loose constraints on how to achieve the goal. As a result, we observed several ways the chatbot adapted to the participant during the inoculation process. \new{The chatbot successfully dealt with various attitudes towards the task and the subject, and unexpected directions during the conversation.}

For example, it often gave options to the user so they can choose how to approach the topic \textit{``Can you think of any research, studies, or personal experiences that might disprove this idea?''} (Chatbot to P6, discussing Exercise and M.H.) 
P6 shared a personal experience, which the chatbot used to recommend relevant literature searches to find scientific evidence, thereby equipping them with refutational defenses beyond anecdotal evidence. 

\new{We observed that participants typically disagreed with the misinformation presented by the chatbot. Recognising the logical fallacy or factual error, they corrected the chatbot or shared their own differing opinion. The chatbot was designed to recognise this as a desirable behaviour; it proceeded to reinforce the participants' beliefs by asking them to find evidence and articulate why they disagree with the misinformation presented.}

Further, we found example of the chatbot adjusting its tone and general approach based on the participant's needs. For instance, it normally promotes independent research, but when P31 asked \textit{``Do you have some scientific based info for me?''}, signifying they might not have any experience in finding scientific information online, the chatbot responded with a scientific database and keywords to search for. 

The chatbot exhibited consistent standards, but some flexibility about what is acceptable as a refutation. Parcipants with responses such as \textit{``i think it is very important to protect nature''} (P53) were asked to begin researching and back up their claims. Those who were able to immediately critically question the misconceptions like P54 \textit{``does it depend on the material of the brush, and how long and vigorous the brushing is?''} were given less guidance and were rather given follow-up questions: \textit{``How could you investigate whether daily brushing truly harms enamel, using resources like online searches or dental expert advice?''}. P54 was reluctant to do research but efficiently used common sense and logic to refute the misconceptions. The therefore no longer asked for resources but allowed the participant to rely on their own reasoning as long as it was factually and logically sound.

In one case, the chatbot dealt with an unexpected event and took it as an opportunity to educate the participant: 
When Forty asked the participant to leave the chat, they responded \textit{``No, i'd like to talk to you more''}. What followed was an educational exchange regarding dentist visits and how to find reliable information about their importance.

\paragraph{Fostering independent thinking}
\old{A core mechanism of successful inoculation was the chatbot’s ability to foster independent thinking. }
\new{We identified the chatbot’s ability to foster independent thinking to be a strength of the system and a potential contributor to the inoculation outcomes.} Unlike traditional inoculation that pre-emptively provides counter-arguments, the chatbot deliberately withheld authoritative information. Instead, it used targeted prompts to guide users toward self-directed inquiry. This process was a critical transition point that transformed the user from a passive recipient of information into an active knowledge-seeker. Conversations that demonstrated this scaffolding often reflected a deeper, more self-constructed understanding. For instance, a user's simple opinion, \textit{``I think it depends on which toothpaste you use,''} was met not with a direct answer, but a query: \textit{``How could you verify which toothpastes are abrasive... Could you investigate and share your thoughts?''} The user's subsequent response, \textit{``Some toothpastes contain abrasive ingredients to aid in whitening the teeth but could ultimately damage tooth enamel''} (P1), demonstrates a successful activation of agency in the participant. Even when P35 asked the chatbot to share resources, the chatbot pushed back and explained \textit{``having you search for resources strengthens your ability to counter misinformation on your own. Try looking up articles or studies that examine the psychological effects of physical activity.''} Similarly, in a discussion on binge-drinking, the chatbot's prompt to \textit{``find any credible sources that might challenge this view?''} elicited a robust counter-argument from the participant: \textit{``Surprisingly it does, repeated episodes of binge-drinking contributes to liver and other chronic diseases as well as increases the risk of several types of cancer...''} (P2). 

\paragraph{Building trust through partnership}
Our analysis revealed that the chatbot's effectiveness was not predicated on it being an authoritative information source, but on its role as a supportive conversational partner. The systematic use of rapport building was a critical design element. Instead of dictating a correct belief, the chatbot validated the user's effort and findings, fostering a sense of shared inquiry. This shift from a hierarchical, expert-to-novice dynamic to a collaborative, peer-to-peer one was instrumental in building trust. This trust, in turn, appeared to lower the user's cognitive defenses, making them more receptive to the inoculation process and more likely to engage in \textbf{reflection} on their own learning. This finding suggests that for sensitive  topics, designing for trust and collaboration is as important as the factual content itself. In response to a user’s research, the chatbot replies: \textit{"Great research! You've shown that some toothpastes can be abrasive, but still, regular brushing is important"} (Chatbot to P1). A participant’s reflection on their own learning at the end of the conversation, \textit{"I have learned a lot, some things I had never thought of."} (P8), suggests that the supportive dialogue led to a genuine shift in their understanding.

\paragraph{Interactional friction as a barrier to engagement}
While the design successfully activated cognitive processes, the analysis also uncovered critical barriers that inhibited the inoculation process. These failures were rooted in interactional friction, such as disruptions in the conversational flow that impede the user's ability or motivation to engage. This finding highlights a critical design implication: even a theoretically sound persuasive strategy can fail if the interaction itself is not robust and seamless. For example, a P1's report of a technical issue, “I'm unable to press continue”, demonstrates friction that immediately halted the process. In some cases, a participant’s brief, seemingly disengaged responses like \textit{“ok”} (P14), effectively prevented the chatbot from continuing its scaffolding and rapport-building efforts.

\paragraph{Summary}

Our thematic analysis directly addressed RQ3 by providing a process-oriented understanding of the factors that promoted and inhibited conversational inoculation during our experiment. We found that conversational inoculation effectiveness is deeply intertwined with the interaction design. 
The factors that promoted the process were tied to a successful conversational user experience. Specifically, a design that fosters independent thinking and metacognitive reflection by prompting users to find and evaluate their own evidence.
Further, a chatbot persona that builds trust through partnership rather than authority was identified as a key mechanism for encouraging this engagement. The primary factors that inhibited conversational inoculation were forms of interactional friction. Whether due to technical glitches or user disengagement, it highlighted the fragility of our conversational method and the necessity of robust interaction design for the theory to be effective in practice.

\subsubsection{Linguistic analysis of chatbot conversations}
 
LIWC assigns scores to input texts in many categories, based on the count of words related to a certain subject or tone. 
First, we used these scores to find linguistic features explaining inoculation outcomes by conducting a Spearman's correlation analysis of the LIWC scores and the post-attack certainty score differences. 
Then, to place the LIWC scores of our chatbot conversations in context, we compared them to reference values. We identified the LMSYS-Chat-1M dataset \cite{lmsys-chat-1m} as a potential reference point, which contains conversations with a variety of models for many purposes.  
As determined using Shapiro-Wilk tests on the 118 variables output by LIWC, we find deviations from normality. Therefore we used statistical tests that do not assume normal distributions.

\paragraph{Correlations with inoculation outcome}
\label{sec:liwc_correlation}

To identify linguistic features that might predict the success of the inoculation, we conducted a correlation analysis between LIWC category scores from the chatbot conversations and Inoculation Effectiveness. We repeated all LIWC correlation analysis three times, once only considering participant messages, once with chatbot messages and once all together. We applied Bonferroni correction to the p-values to adjust for multiple comparisons. 


Our analysis revealed no statistically significant correlation between LIWC variables and overall inoculation effectiveness. 
Due to the potential negative effect of the variety of topics on the correlation detection, we followed-up with tests for individual topics. 
Across the four topics only Binge Drinking showed any correlation, where the LIWC "Achieve" variable was strongly associated with inoculation effectiveness (Spearman's $\rho$ = 0.92, $p = .008$) when analysing the overall conversations. 
The Achieve variable stands for the Achievement tone, which is a type of Drive, which is a type of Psychological Process in LIWC's taxonomy. 
It includes words such as ``work'', ``better'', ``best'', ``working'', ``try'', ``goal'' and ``win''.
This finding suggests that inoculation of specific subjects may benefit from using a specific language, tone or subject framing. 

\paragraph{Comparison to reference chatbot conversations}

To contextualise the nature of the conversations within our experiment, we compared their linguistic profiles to sample ($N = 10,000$) from real-world user-LLM conversations from the LMSYS-Chat-1M dataset \cite{lmsys-chat-1m} collected in 2023. It comprises of mostly English conversations with chatbots such as Vicuna, Koala, Alpaca, Llama and Claude. Among the most common discussion topics are software (coding, troubleshooting), geography and travel, text processing, business strategies and roleplaying. We performed a t-test to identify LIWC categories where our conversations differed significantly ($p < .01$) and with a medium-to-large \cite{Cohen_2009} effect size (\textit{Cohen’s d}~$> 0.5$) from this reference dataset.

As seen in Table \ref{tab:liwc_results_selected}, there were significant differences in the participants' messages. Our participants demonstrated lower Clout (associated with relative social status, confidence, or leadership) and Analytic language (associated with logical, formal language, good reasoning skills), suggesting a less formal and more conversational tone compared to the reference corpus.

Forty the chatbot used more words related to Curiosity, reflecting its designed role as an engaging conversational partner. Conversely, the bot's messages were less Analytic. The chatbot also used more Authentic (associated with perceived honesty, genuineness) language and its messages contained more Cognitive Processes (such as ``but'', ``not'', ``if''), indicating they may have been more complex than the average chatbot message. 

The complete conversations (human and chatbot messages together) showed a higher frequency words related to the topic of Health verifying that the discussions had the intended theme. Consistent with the bot-only findings, the complete conversations were also less Analytic. Language related to Conflict indicates disagreement between the participant and the chatbot, verifying that a core component The significant difference in word count (WC) further demonstrates that the core component of Cognitive Inoculation, refutation, was succesfully elicited in the conversations. conversations in our study were more extensive than those in the reference corpus.

\begin{table*}[t]
\centering
\footnotesize
\caption{Selected LIWC variable differences filtered by effect size \textit{Cohen’s d} $> 0.5$ and significance $p < .01$, grouped by level of comparison. See the complete list in Appendix K. Reference values are from \cite{lmsys-chat-1m}.}
\label{tab:liwc_results_selected}
\begin{tabular}{lrrrrrrr}
\toprule
\textbf{LIWC Variable} & \textbf{M (MindFort)} & \textbf{SD (MindFort)} & \textbf{M (Reference)} & \textbf{SD (Reference)} & \textbf{t} & $\textbf{df}$ & \textbf{d} \\
\midrule
\addlinespace
\multicolumn{8}{l}{\textbf{Human messages only}} \\

Clout & 31.26 & 19.32 & 53.79 & 35.14 & -9.00 & 10059 & -0.64 \\
Analytic & 50.30 & 24.70 & 69.52 & 32.91 & -6.04 & 10059 & -0.58 \\

\addlinespace
\multicolumn{8}{l}{\textbf{Chatbot messages only}} \\

Curiosity & 2.27 & 0.96 & 0.49 & 1.14 & 14.39 & 10021 & 1.56 \\
Authentic & 72.75 & 15.70 & 33.58 & 32.63 & 19.22 & 10021 & 1.20 \\
Analytic & 34.94 & 10.14 & 70.20 & 31.33 & -26.36 & 10021 & -1.13 \\
Cognitive Processes & 20.59 & 2.74 & 11.59 & 8.04 & 25.01 & 10021 & 1.12 \\

\addlinespace
\multicolumn{8}{l}{\textbf{Complete conversations}} \\

Health & 3.20 & 2.96 & 0.57 & 1.67 & 6.93 & 10059 & 1.57 \\
Analytic & 42.48 & 10.92 & 72.28 & 28.22 & -20.88 & 10059 & -1.06 \\
Authentic & 59.47 & 16.24 & 32.13 & 29.65 & 13.01 & 10059 & 0.92 \\
Tone & 74.26 & 22.84 & 46.45 & 31.14 & 9.45 & 10059 & 0.89 \\
Cognition & 27.33 & 3.06 & 20.70 & 8.20 & 16.59 & 10059 & 0.81 \\
Conflict & 0.57 & 0.40 & 0.18 & 0.61 & 7.43 & 10059 & 0.63 \\
Emotion & 1.97 & 1.15 & 1.01 & 1.61 & 6.53 & 10059 & 0.60 \\
WC & 632.26 & 419.93 & 337.11 & 570.53 & 5.46 & 10059 & 0.52 \\
\bottomrule
\end{tabular}
\end{table*}

\section{Discussion}

Our study introduces and investigates Conversational Inoculation, a method for building resistance to misinformation through a dialogue-based, LLM-powered agent. 
The results from our experiment with the MindFort prototype provide initial evidence for the viability and potential of this approach.
In this section, we discuss our findings in the context of our research questions and consider the broader implications for the field of human-computer interaction

\subsection{Effectiveness of Conversational Inoculation (RQ1)}
Our results provide partial support for the effectiveness of Conversational Inoculation, as implemented in MindFort.
In a direct between-condition comparison, Chatbot outperformed Control but did not differ statistically significantly from Reading or Writing.
This establishes Conversational Inoculation not necessarily as more effective, but certainly as a valid method alternative already now, with many different technological affordances in the future.
Further, when adjusting for each participant’s baseline susceptibility (their certainty change in the Control condition, after a misinformation attack), Chatbot yielded greater inoculation effectiveness than Reading and Writing.

A possible explanation is the cognitive effort required to engage in a conversation and construct counter-arguments.
In traditional active refutation, participants are explicitly tasked with generating arguments.
With MindFort, this process is scaffolded by the chatbot, which might reduce the cognitive load but potentially also the depth of processing for some individuals.
Future experiments could benefit from think-aloud or process-tracing methods to understand the mental processes involved in Conversational Inoculation.

The effectiveness of Conversational Inoculation varied across topics (see \autoref{sec:topic_differences}). While a study like ours cannot fully explain these differences, the results suggest that some topics may require tailored approaches.
In particular, prompt design that accounts for contextual factors relevant to the topic at hand could be an important direction for future research.

\subsection{User Experiences with Conversational Inoculation (RQ2)}

Internal Motivation Inventory (IMI) and open feedback from participants regarding their experience with MindFort address RQ2.
The Chatbot condition scored similarly high on all IMI subscales as the other conditions, indicating that people would have a similar experience using a system implementing Conversational Inoculation as the other methods implemented within MindFort.
Yet, future work could refine the design to isolate participants' responses to the treatment component more directly to understand granular differences in the inoculation methods rather than the full inoculation process.

As shown in \autoref{sec:feedback}, qualitative feedback shared by the participants after each lesson and the debriefing verified these findings, as both MindFort and the Chatbot itself were found to be compelling in various ways.
This is important, as the success of any educational intervention, especially those designed to build long-term resilience, hinges on user engagement and sustained use. 
User engagement can also lead to fatigue \cite{Tang_Sun_Nie_Li_Sergeeva_LC_2025} if the inoculation treatment has a high cognitive load or takes too long. Based on our preliminary IMI and qualitative feedback results, Conversational Inoculation, being relatively flexible in discussion subjects and timing, is not inherently too fatiguing to users.

The user experience can be attributed to our design of the overall web interface and the chatbot persona itself. 
By establishing the chatbot as a credible and reliable expert and the inoculation task as a polished and fun learning activity, we likely primed participants to be more receptive, as suggested in~\cite{Banas_Rains_2010}.
The conversational turn-taking also fostered a sense of agency and perceived competence, as participants felt they were actively contributing to the process rather than just being lectured.
Negative feedback about the chatbot's undesirable behaviours shows a clear path to improve the method in the future by addressing the interactional friction and conversational rigidity that participants identified. 
Future prototypes should, for example, use few-shot prompting that includes giving the chatbot examples of good conversations (as proven by the measured inoculation outcome) or potentially fine-tuning a model for Conversational Inoculation.

\subsection{Factors Influencing Conversational Inoculation (RQ3)}

RQ3 investigated the underlying mechanisms that promote or inhibit the effectiveness of Conversational Inoculation.
By combining qualitative thematic analysis of the chatbot dialogues with quantitative linguistic analysis, we identified several key factors related to the user's cognitive engagement and the quality of the interaction itself.

Our thematic analysis revealed that successful inoculation was often predicated on the chatbot's ability to foster independent thinking.
Rather than simply providing counter-arguments, the agent prompted users to actively seek out and evaluate evidence on their own.
This scaffolding of the refutation process transformed users from passive recipients into active participants, deepening their cognitive engagement.
This collaborative dynamic appeared to lower users' resistance and encourage reflection, making them more receptive to the inoculation process.
The LIWC analysis also confirmed that the experimental dialogues were significantly more analytical and cognitively demanding for both the user and the bot compared to typical chatbot interactions. This reinforces the idea that this heightened cognitive engagement is a core component of the conversational inoculation process. 

Conversely, the primary inhibitor of Conversational Inoculation was interactional friction.
This included technical issues that disrupted the conversational flow as well as user disengagement which may be less of an issue when user's engage with the system in an unpaid setting or because of their intrinsic motivation.
When this friction occurred, it prevented the chatbot from effectively guiding the user through the necessary cognitive steps, thereby undermining the entire process. 

The LIWC analysis provided further, albeit nuanced, insights.
While no single linguistic marker correlated with inoculation success across all topics, we identified a positive correlation between the use of achievement-oriented language and inoculation effectiveness for the specific topic of binge-drinking.
This suggests that the optimal conversational tone and style may be topic-dependent, requiring the agent to adapt its language to align with the specific psychological drivers of a given attitude.
We hypothesise that in this specific case, framing binge-drinking as a struggle or progress (where positive achievements can be made) evoked emotions in the participants that had a significant effect on their opinions or their resistance to subsequent counterattitudinal messages.  

Taken together, these findings suggest that the success of Conversational Inoculation hinges on the delivery of information as well as on the interaction design, much in line with findings in~\cite{Gong2025276}.
An effective system must not only present counter-arguments but also actively foster the user's independent research and learning skills, build a trusting rapport, and maintain a seamless, engaging conversational flow tailored to the subject at hand.

\subsection{Broader Implications and Future Work}

Our study reveals a nuanced value proposition for Conversational Inoculation, particularly when considering the contrasting outcomes between conversational and other treatments.
One of its core values lies in its potential for adaptive conversations. Conversations can adapt to refutational styles, it can be more educative and guided if the user needs, or can become engaging debates, as observed in the conversations in \autoref{sec:qualchatbotconv}.
This is a significant departure even from interactive technological inoculation methods \cite{Jeon_Kim_Xiong_Lee_Han_2021, Basol_Roozenbeek_VanDerLinden_2020} as they often still follow rigid, pre-defined learning paths. However, our participants reported high engagement and a feeling of being entertained and educated by the chatbot. 
Further, IMI results suggest that Conversational Inoculation does not lower user motivation to use and learn with the system.
In light of the outcomes of this study, we discuss the broader implications of Conversational Inoculation and identify future directions.

\paragraph{Personalised and dynamic inoculation strategies}

A critical area for future development lies in tailoring inoculation strategies to individual users and contexts. 
The current implementation is shown to adapt to the participant in different ways. However, we hypothesise that rich background information about them to plan a personalised inoculation approach further benefits the inoculation outcome. 
This future potential is the second core value we identified.
Moving forward, chatbots should incorporate explicit mechanisms for dynamically adjusting inoculation content, delivery style, and interaction complexity based on real-time user responses and inferred psychological states. 
This could involve adapting the level of challenge presented, the types of persuasive arguments countered, and the emotional tone of the interaction to maximise engagement and learning. 
Further, future systems could leverage user profile data, past interaction history, and even facial expression analysis to personalise the inoculation experience and increase its impact.

\paragraph{Exploring other forms of conversation}

While our study implemented a single-bot, single-user dialogue, future work should explore the rich design space of conversational agent configurations. The introduction of an interactive agent raises new questions beyond simple prompting. As highlighted by \citet{Matusitz_Breen_2010}, the perceived personality and credibility of the inoculator are crucial for success. Future research could investigate a variety of chatbot personas and the roles they play. 
For instance, a bot could act as a skeptical learner, a collaborative research partner, or a more assertive educator. 
Each role creates a distinct conversational flow and potentially altering the ideal level of user independence required for the task. As outlined in the previous section, it may even be beneficial to automatically make the decision of this configuration based on knowledge about the user. 

Further, a key limitation we observed in our single-bot approach was the tension between its two functions: acting as a threat and acting as a helpful guide. The bot's apparent motivation to help the user often decreased the perceived threat of its counter-attitudinal arguments. This challenge could be addressed by distributing these roles across multiple agents. For example, a "multi-bot inoculation" system could feature one chatbot dedicated to attacking the user's beliefs, while a second, neutral "facilitator" bot assists the user in researching and forming counter-arguments. With a strong foundation of work on multi-agent conversational LLM systems already available \cite{geng2025beyond, Becker_2024,Ye_Liu_Wu_Pang_Yin_Bai_Chen_2025, Yan_Zhang_Zhang_Zhang_Zhou_Miao_Li_2025}, researchers can begin to experiment with various configurations of agent roles and personas, such as facilitator-peer or expert-novice dynamics \cite{geng2025beyond}.

\paragraph{Multi-bot inoculation}

We acknowledge that the chatbot assumed two contradictory roles. It aimed to simulate threat by presenting itself as a believer of the misconceptions, but at the same time maintaining an educational, friendly and accommodating approach. Therefore, we think that further potential of the conversational inoculation method may be unlocked if the two roles are taken on by not one but two or more chatbots. For example, one chatbot attacking the subject's health attitudes, and another chatbot, assisting the subject to engage in critical thinking, individual research and disprove the first bot's claims. There is now plenty of work on multi-agent conversational LLM systems available \cite{geng2025beyond, Becker_2024,Ye_Liu_Wu_Pang_Yin_Bai_Chen_2025, Yan_Zhang_Zhang_Zhang_Zhou_Miao_Li_2025} to support the design of this conversational system, and future research should explore variously configured multi-bot conversations and new roles (e.g. a facilitator and peers \cite{geng2025beyond}) among them.

\paragraph{Thinking beyond just persuasion}

Researchers are now investigating whether cognitive inoculation can be applied beyond attitudes, such as self-efficacy or autonomous motivation~\cite{Compton_2024}
These are constructs that may receive attacks similar to attitudes \cite{bandura77,reeve09}, reducing one's self-efficacy or autonomous motivation, potentially e.g. affecting one's health and wellbeing. 
In future work, conversational systems similar to MindFort should investigate inoculating other cognitive constructs, as those might benefit from the conversational interfaces in different ways.
For example, a conversational inoculation treatment can be designed to deliver preemptive messages to bolster an individual's sense of self-efficacy in high-pressure environments such as academia.
These settings could further benefit from the bot's ability to deliver personalised and dynamic treatment, as the struggles and types of attacks that students and staff may receive on their self-efficacy and individual needs and goals can vary widely. 

Further, as discussed in \autoref{sec:limitations}, inoculating against phenomena promoting the dissemination of misinformation may lead to an overall larger positive impact. 

\subsection{Limitations}
\label{sec:limitations}
We acknowledge several limitations in our work. 

\old{The experiment was conducted using Prolific, a human subject pool. 
Although Prolific has been shown to provide high-quality research data for human subject experiments, the study experienced a high dropout rate due to the long task. 
This attrition partially affected the success of our counterbalancing, as several participants dropped out mid-experiment.}
First, it was difficult to introduce a long interval between the inoculation treatment and the persuasive attack. 
Longer intervals are typically associated with better inoculation outcomes, as they allow for greater knowledge consolidation~\cite{godbold2000}. 
Nevertheless, the results highlight differences between the three treatments and the baseline, and longer studies in the future can address these challenges.

Second, our study was limited to health, wellbeing and nature topics which is a a small sample of all possible topics. 
This leads to sub-optimal generalizability of the findings to other topics as discussed in \autoref{sec:topic_differences}. 

\new{Third, the} small sample size also prevented topic-specific analysis, making it difficult to generalize our findings to other topics where personal involvement might be higher. 
Even though cognitive inoculation is often criticised for its narrow focus on specific topics \cite{Basol_Roozenbeek_VanDerLinden_2020}, we chose to focus mainly on wellbeing-related topics to keep the work manageable and scoped. 
In the future, we may overcome this by also inoculating against the underlying phenomena such as echo chambers, the Dunning-Kruger Effect\cite{dunningkruger}, or cognitive biases such as familiarity bias, confirmation bias, or availability bias. 
Addressing the mechanisms that produce misinformation, such as inoculating against the echo chamber effect \cite{Jeon_Kim_Xiong_Lee_Han_2021}, rather than targeting only specific topics, may foster broader critical thinking skills and strengthen resistance across a wider range of attacks.

\new{Fourth, regarding the Control condition, we implemented a passive baseline (absence of treatment). This design choice was guided by the principle of avoiding ``overcontrol bias'' \cite{cinelli2024crash}; since the conversational interaction is a mediator of the intended effect, controlling for the medium (e.g., via a baseline chatbot) would block a key mechanism we aimed to estimate. Consequently, our results reflect the \textit{total} causal effect of the intervention. The trade-off of this approach is that we cannot strictly isolate the specific effects of the refutation content from the general engagement effects of interacting with a chatbot. Future work could explicitly investigate these component effects by including an active control, such as a non-argumentative ``descriptive'' chatbot.}

\new{Finally, our linguistic analysis did not identify significant linguistic markers of overall inoculation effectiveness. We acknowledge that this may be due to the strict Bonferroni correction applied to the high number of LIWC variables tested, which minimizes false positives but increases the risk of Type II errors (missing actual effects). Future research might benefit from a more targeted analysis focusing on a narrower, theory-driven subset of linguistic features.}

\section{Conclusion}

This work explored Conversational Inoculation, implementing it in a Web-based system for cognitive inoculation.
We demonstrate the feasibility of this approach and offer preliminary evidence supporting its effectiveness in safeguarding against persuasive attacks. 
Conversational Inoculation was found to be an equally valid method compared to the traditional methods of reading and writing.
When controlling for baseline susceptibility to misinformation, it outperformed them both in our experiment.
The emergence of large language models makes it feasible to implement versatile, online dialogue-based interventions.
Our work provides a timely case study on combating misinformation and positions Conversational Inoculation as a promising research direction in HCI.


\begin{acks}
This research is partly funded by the Strategic Research Council (SRC), established within the Research Council of Finland (Grants 356128, 335625, 335729), and Research Fellowship funding by the Research Council of Finland (Grants 356128, 349637 and 353790). The research was also partly supported by JSPS Bilateral Collaboration between Japan and Finland (Grant Number: JPJSBP120232701). 
This work was partly supported by JST ASPIRE for Top Scientists (Grant Number JPMJAP2405).
\end{acks}

\bibliographystyle{ACM-Reference-Format}
\bibliography{main}

\newpage



\end{document}